\documentclass[10pt,twocolumn]{article}

\usepackage{amsxtra,amssymb,amsthm,amsmath,latexsym}

\usepackage{graphicx}

\newcommand{\bee}{\begin{equation*}}
\newcommand{\eee}{\end{equation*}}
\newcommand{\be}{\begin{equation}}
\newcommand{\ee}{\end{equation}}

\title{Phase imaging from defocus information in a light field}
\author{Q. Tyrell Davis}
\date{}
\begin{document}

\twocolumn[
  \begin{@twocolumnfalse}
\maketitle
\begin{abstract}
Optical microscopy is without a doubt an essential component of life science research, but many objects of interest in biology are transparent. Chemical or immunological dyes, which can often be toxic, fluorescent transgenes, which require a protocol for transformation of exogenous DNA, and phase contrast, which inextricably combines phase information with amplitude, all serve to increase the contrast of transparent objects. Transparent objects of non-uniform refractive index and/or thickness are phase objects and phase can be extracted from defocus information above and below the object. The light microscopist will recognize this in that object contrast can vary as the stage is moved around an object's focal plane. This phase information can be quantitatively retrieved from a set of defocused images, but this has the disadvantage of decreasing temporal resolution, as the microscope stage has to be moved between image captures. By incorporating additional optics in the form of an array of pinholes or microlenses,	 the direction and position of all the rays can be recorded in a light field, a 4D function. By recording the light field, multiple planes of focus can be investigated, allowing for the capture of defocus information in real time. Here I discuss and investigate the use of the defocus information in a light field to obtain the transport of intensity equation, and use the fast Fourier transform method to solve this equation for phase.

%Light field imaging with a microscope allows the imaging of parallax and exploration/recording of multiple focal planes in real time. Quantitative phase imaging can increase contrast of transparent specimens, a category to which many microscopic objects belong, under the microscope, as well as providing quantitative measurement of the product of refractive index and thickness. When thickness or refractive index is known, the either can be measured with great sensitivity.Phase imaging can be computed from defocused images of an object, this information is intrinsically captured in the recording of a light field. By combining these two techniques, many new capabilities may be realized. 

%By capturing the entire light field any application of quantitative phase imaging can be pushed into real-time, allowing for the observation and measurement of dynamic events, and at relatively cheap cost. Additionally, novel applications are likely to emerge. I anticipate that light field quantitative phase imaging might alleviate preparation requirements in clinical histology: increased contrast from phase calculation may relieve the need for exogenous dyes, while the multiple focal plane capture of light field imaging could add spatial information. The interactivity of the light field could allow for clinical diagnoses to occur off-site, relieving the technician of the burden of capturing precisely the correct focus.
\end{abstract}
%t{\em Key words:}  phase, light field
  \end{@twocolumnfalse}
  ]

\section*{Introduction}
Biological samples of interest to the microscopist are often phase objects, that is, they are largely transparent and offer low contrast when viewed under light illumination. Although transparent, they do perturb the phase of transmitted illumination via refractive index variation. This phase variation can be used to qualitatively increase contrast as in  Zernike phase contrast microscopy \cite{zernike} through interferometric means. In the past twenty years or so, phase has also been obtained quantitatively through a variety of means, one of the easiest being the transport of intensity equation\cite{streibl1984} (TIE, equation \ref{TIE}) which describes the change in intensity per change in axial distance from a given focal plane. The simplest way to obtain the transport of intensity equation is to record three or more images above, at, and below the focal plane in which the object of interest resides \cite{streibl1984}. Physically manipulating the microscope stage for defocus decreases temporal resolution, so it is desirable to obtain defocus information in real time. Waller et al demonstrated that this can be accomplished by separating the color channels in a three color image, taking advantage of the chromatic aberration in microscope objectives, specifically those that are either uncorrected for chromatic aberration or achromats, corrected for two colors\cite{waller2010}. Treating three color channels as three separate images at slightly different focus allows for the capture of quantitative phase in real time, allowing phase observation of dynamic events and video capture. The subject of study must be uncolored, making this method incompatible with say, fluorescence microscopy. Also, phase computation accuracy can be improved by incorporating many defocused images\cite{soto2007}, whereas recovery based on chromatic aberration is limited to the color channels on a camera sensor. 

Light field microscopy can be achieved by inserting a microlens array with desired characteristics into the intermediate image plane of a microscope, and allows for the computation of multiple focal planes in a single capture \cite{levoy2006}. This is strikingly amenable to real-time phase computation by solution of the TIE, and has the advantage that many planes can be used to estimate the axial derivative of irradiance. To demonstrate the capability to compute quantitative phase from defocus information contained in a light field, I generated a focal stack of images using a sample light field from the Stanford Light Field Microscope Project website (http://graphics.stanford.edu/projects/lfmicroscope/) and built a MATLAB algorithm for solving for the associated TIE according to the fast Fourier transform methodas outlined in \cite{popescu2011}

\section*{Method}
To demonstrate the feasibility of calculating phase information from defocus information contained in a light field, I utilized a sample light field available from the Stanford light field microscopy project\cite{lfmicroproj}. Using the light field viewing software LFDisplay (available from the same source) I captured an image focused at zero as well as ten microns above and below utilizing screenshot capture. For experimental applications outside of this concept note, data would be output directly from the light field manipulation software rather than as a copy of the monitor display. 
%comment on uniformity of illumination intensity

The intensity difference of the two defocused images contains phase information retrievable via the transport of intensity equation\cite{streibl1984}. The transport of intensity equation\cite{popescu2011}$^{p. 222, eq. 12.2}$:

\begin{equation}
k_0 \frac{\delta I(x,y)}{\delta z} = -\nabla \cdot [I(x,y) \nabla \phi(x,y)]
\label{TIE}
\end{equation}

Where $k_0$ is the wavenumber $\frac{2\pi }{ \lambda}$, $I(x,y)$ is the intensity, $\frac{\delta I(x,y)}{\delta z}$ is the difference in intensity for axial increments in z, and $ \phi(x,y)$ is the phase. $\frac{\delta I(x,y)}{\delta z}$ is obtained empirically from a minimum of two defocused images\cite{popescu2011}$^{p. 222, eq. 12.4}$.

\begin{equation}
\frac{\delta I(x,y)}{\delta z} = \frac{1}{2 \Delta z} [I(x,y, \Delta z) - I(x,y -\Delta z)]
\end{equation}

\begin{quotation}
$ = g(x,y)$
 \end{quotation}

Where $\Delta z$ is the distance amount of defocus. To recover the phase from the difference image $g(x,y)$ we assume, albeit counter-intuitively, that the image intensity on the right in Equation \ref{TIE} is constant. This gives \cite{popescu2011}$^{p. 222, eq. 12.3}$:

\begin{equation}
g(x,y) = - \frac{I_0}{k_0} \nabla^2 \phi (x,y)
\label{gxy}
\end{equation}

To solve the Laplacian ($\nabla^2$) in Equation \ref{gxy}, we take the fast Fourier transform (FFT) of $g(x,y)$ to yield $G(k_x,k_y)$. In my MATLAB script, this is obtained with the 2-dimensional FFT command, fft2(). Phase in the frequency domain, $\Phi(k_x,k_y)$ is obtained from the following \cite{popescu2011}$^{p. 223, eq. 12.5}$:

\begin{equation}
\Phi(x,y) = \frac{k_0}{I_0} \cdot \frac{G(k_x,k_y) }{(k_x^2 + k_y^2)}
\label{PHI}
\end{equation}

$I_0$ is the intensity distribution at the focal plane, assumed to be uniform. This value is obtained by taking the average of the in-focus image intensity. To convert the phase $\Phi(x,y)$ in Equation \ref{PHI} to the spatial domain, we take the 2-dimensional inverse FFT.

To compute phase from the defocus stack obtained with LFDisplay I wrote a MATLAB function, which uses the fast Fourier transform to recover phase information from the intensity derivative from a three image focus stack. The inputs are images above, below and at the focal plane of interest and the defocus distance between images, and the function returns a phase map. This function is included as an appendix to this note.

\section*{Example}
\begin{figure*}[hp!]
\begin{center}
\includegraphics[keepaspectratio, scale = 0.36]{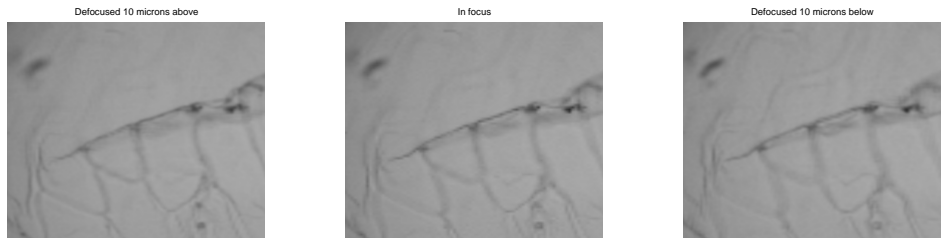}
\caption{{\bfseries Three 2-dimensional images separated by 10 microns of focus were retrieved from the light field. These will be used to compute the phase using the transport of intensity equation (TIE). Original LF image source: http://graphics.stanford.edu/projects/lfmicroscope/2006.html}}
\label{onionstack}
\end{center}
\end{figure*}

\begin{figure}[hp!]
\begin{center}`
\includegraphics[keepaspectratio, scale = 0.35]{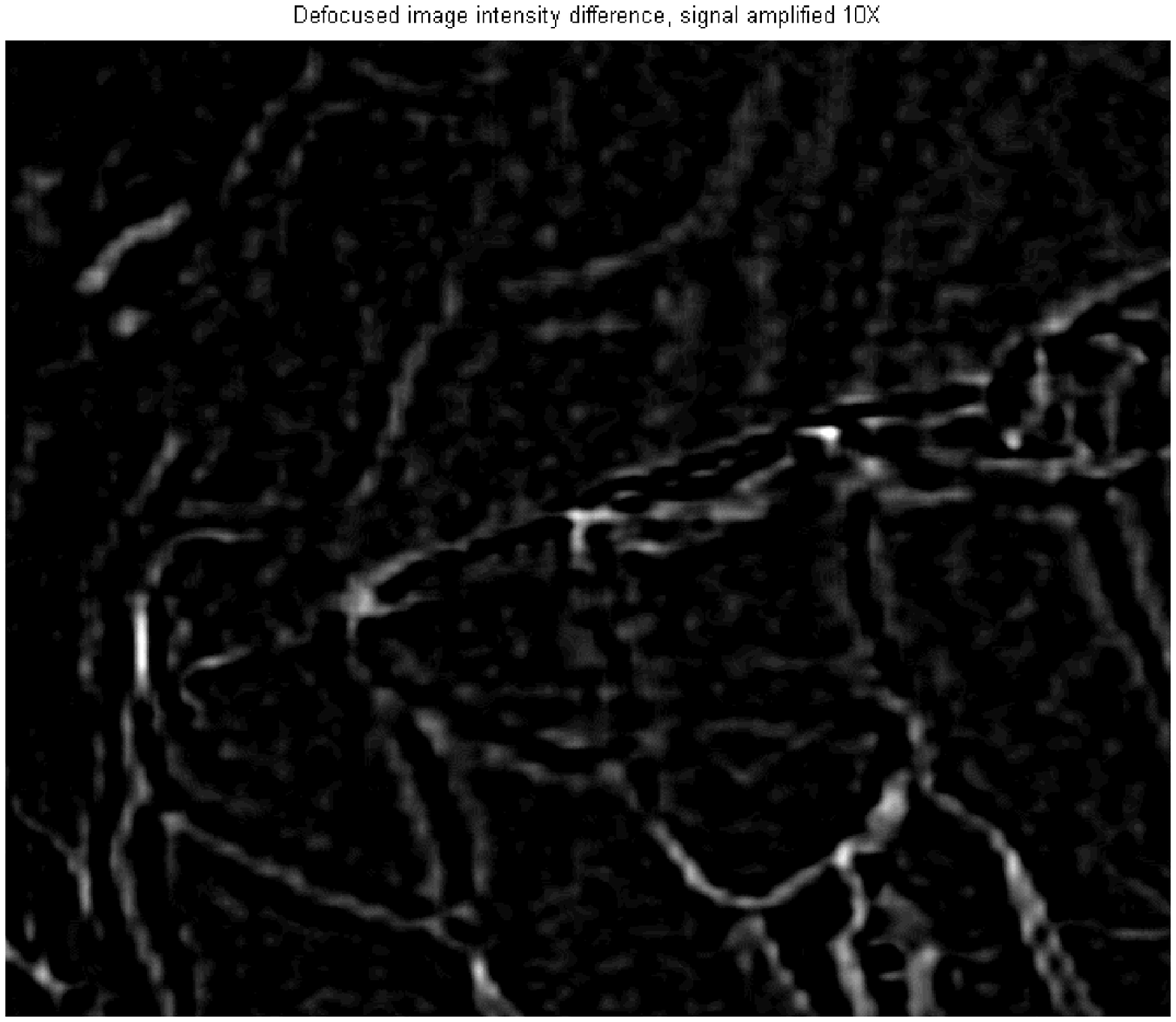}
\includegraphics[keepaspectratio, scale = 0.2]{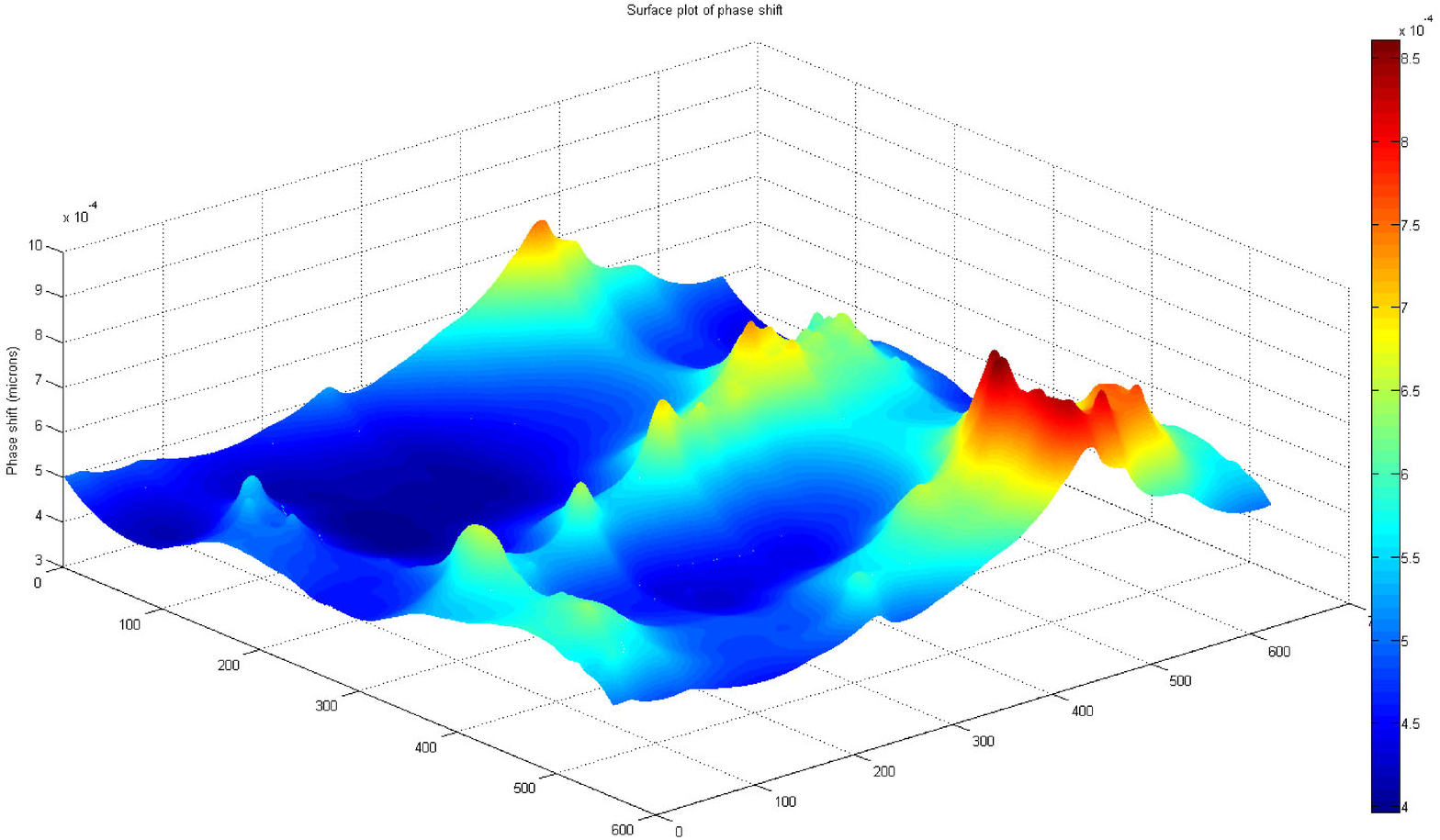}
\caption{{\bfseries The defocus difference image (top, signal amplified 10X) is used to determine the function g(x,y) (Equation \ref{gxy}), which is used to compute the phase map via Equation \ref{PHI} (bottom). The fold in the onion skin is readily discernible in the phase map, as well as certain details of the nodules in the vascular network of the plant. }}
\label{phase}
\end{center}
\end{figure}

A sample light field was used to test the feasibility of recovering phase. The subject was an onion skin taken with a 20X magnification, 0.5 numerical aperture objective, imaged onto a 24X36 mm array of square 125 micron pitch, f/20 microlenses, obtained from \cite{lfmicroproj}. Three images separated by 10 microns of focus were used to calculate phase (Figure \ref{onionstack}). The difference image representing $\frac{\delta I(x,y)}{\delta z}$ (amplified by a factor of 10 for viewing) and the recovered phase map are the contents of Figure \ref{phase}.

\section*{Conclusions}
Any application of quantitative phase imaging could benefit by the use of light field imaging to push the technique into real time, for a relatively inexpensive cost of one or more interchangeable microlens arrays tailored to the application. The breadth of data collected in a light field could eliminate the requirements for an expert operator, for example in a rural clinic, because the phase computations and diagnoses can be made at a point spatially and/or temporally removed from the sample and microscope. Applications in manufacturing could include precise quality control of surfaces or structures differing only in thickness or refractive index, for example in laboratory etching of microfluidic devices in glass or silicon. 

There is a tradeoff in light field capture versus traditional 2D microscopy, discussed in \cite{levoy2006} and \cite{levoy2009}: the product of angular and lateral resolution cannot exceed the diffraction limit of the optical apparatus and the wavelengths measured. In simple terms, the pixel area of each microlens image multipled by the total number of microlenses can not exceed the diffraction limit of the microscope. This would limit the observable feature size to a few times larger than the resolution limit of a given optical arrangement, but may be attenuated with further computational tricks and should be a factor in microscope design, in particular in the choice of microlens array characteristics and objective. These considerations are discussed in a tutorial in \cite{lfmicroproj}. I also note that Levoy et al have experimented with light field phase recovery of a homogeneous medium in \cite{levoy2009}. However, their method employed additional optical components in a microlens array and lcd projector screen in the illumination path. 

%Any application in which quantitative phase imaging is advantageous could benefit from light field imaging by pushing the process into real-time and providing a complete data capture in the recorded light field for later analysis. In particular, untreated cells or tissues could be quickly analysed and pathological/healthy tissue distinguished based on their phase signature. By caching a complete view of the tissue or cells in question in the light field, analysis can be performed throughout the light field subsequent to the data capture, and this analysis could be performed off site where second opinions or dedicated expertise are required. In addition to the potential clinical applications, basic research could benefit any time a detailed picture of morphology dynamics is desired. Applications need not be limited to biology and medicine, however, as phase measurements can be used for precise measurements of thickness, for example, in a quality control step for mass produced microfluidic devices or semiconductor wafers. 

%\newpage

\onecolumn
\section*{Appendix}
%\twocolumn[
  %\begin{@twocolumnfalse}

\begin{verbatim}

function [ phase ] = phaseNewWorld(Ia,I0,Ib,z)
%generate a phase map based on the defocused images Ia (above), Ib (below, 
%focused image I0, and the defocused difference (in microns) z 
%   Detailed explanation goes here
%For starters we will split of the green channel and deal with it
%exclusively

if size(size(Ia)) == 3 % check for color/B&W image
	Iag = Ia(1:size(Ia,1),1:size(Ia,2),2);
	Ibg = Ib(1:size(Ia,1),1:size(Ia,2),2);
	I0g = I0(1:size(Ia,1),1:size(Ia,2),2);
else 
	Iag = Ia;
	Ibg = Ib;
	I0g = I0;
end
	
%Iavg is the uniform intensity distribution
Iavg = mean(mean(I0g));

%wavenumber k0 (inverse microns, for green light)
k0g = (2*pi())/(500 * 10^-3);

%gofxy is the data that describe the change in field intensity per axial
%defocus (z). Ironically an assumption of this solution of the TIE is that
%the field intensity is assumed to be constant over z (see p. 223 Popescu 2011,
%QPI of cells and tissues)
gofxy = (Iag - Ibg) / (2*z);

%gofxy contains the phase information [gofxy = -I0/k0 $\nabla^2$ \phi(x,y)]
%To solve the Laplacian we convert gofxy to the frequency domain with the
%2D fft
GOFkxky = fft2(gofxy);

%debugging: show the frequency domain and difference image
%figure(1); subplot(2,1,1); title('difference Ia-Ib'); imshow(Iag-Ibg); 
%subplot(2,1,2); title('FFT magnitude of dI/dz $\delta$'); imshow(abs(GOFkxky));

%Solve for the frequency domain phase:
%Declare memory to hold the fft of phase
PHI= zeros(size(Iag,1),size(Iag,2));
kx = 1;
while kx < size(GOFkxky,1)
    ky = 1;
    while ky < size(GOFkxky,2)
        PHI(kx,ky) = (k0g/Iavg) * GOFkxky(kx,ky) / (kx^2+ky^2);
    ky = ky+1;
    end
    kx = kx+1;
end

phase = ifft2(PHI);
figure(2); subplot(1,3,1); title('defocused above'); imshow(Iag); 
subplot(1,3,2); title('focused image'); imshow(I0g); 
subplot(1,3,3); title('defocused below'); imshow(Ibg);
figure(3);
subplot(1,2,1); title('difference image amplified 10X, dI/dz'); imshow(10*(Iag-Ibg)); 
subplot(1,2,2); title('recovered phase, surface plot'); mesh(abs(phase));
figure(4); title('recovered phase, surface plot'); mesh(abs(phase)); colormap(jet);




end
\end{verbatim}
  %\end{@twocolumnfalse}
  %]

\begin{thebibliography}{00}

\bibitem{lfmicroproj}
Anderson, T., Horowitz, M., Levoy, M., Zhang, Z., Grosenick, L. Stanford Light Field Microscope Project. at $<$http://graphics.stanford.edu/projects/
lfmicroscope$>$

\bibitem{levoy2006}
Levoy, M., Ng, R., Adams, A., Footer, M., Horowitz, M. Light Field Microscopy. ACM Transactions on Graphics 25(3), Proc. SIGGRAPH. 2006.

\bibitem{levoy2009}
Levoy, M., Zhang, Z., Mcdowall., I. Recording and Controlling the 4D light field in a microscope. Journal of Microscopy. 235-2, 144-162. 2009. 

\bibitem{popescu2011}
Popescu, G. Quantitative Phase Imaging of Cells and Tissues. (McGraw-Hill Prof Med/Tech: 2011)

\bibitem{soto2007}
Soto, M., Acost, E. Improved phase imaging from intensity measurements in multiple planes. Applied Optics 46(33), 7978-7981. 2007

\bibitem{streibl1984}
Streibl, N "Phase imaging by the transport equation of intensity". Optics communications (0030-4018), 49 (1), p. 6. 1984.

\bibitem{waller2010}
Waller, L., Kou, S. S., Sheppard, C. J. R. \& Barbastathis, G. Phase from chromatic aberrations. Opt. Express 18, 22817–22825 (2010).

\bibitem{zernike}
Zernike, F. How I Discovered Phase Contrast. Science 121, 345–349 (1955). 

\bibitem{lfguide}
Zhang, Z. A Practical Introduction to Light Field Microscopy. at $<$http://graphics.stanford.edu/software
/LFDisplay/lfmintro/$>$



\end{thebibliography}
\end{document}